\documentclass[letterpaper,11pt]{article}

\usepackage{amsmath,axodraw,amssymb,epsfig}

\begin{document}

\baselineskip=18pt

\newcommand{\eps}{\epsilon}
\newcommand{\pslash}{\!\not\! p}
\newcommand{\m}{\widetilde m_u}
\newcommand{\M}{\widetilde M_u}


\thispagestyle{empty}
\vspace{20pt}
\font\cmss=cmss10 \font\cmsss=cmss10 at 7pt

\begin{flushright}
UAB-FT-567
\end{flushright}

\hfill
\vspace{20pt}

\begin{center}
{\Large \textbf
{The Minimal Composite Higgs Model}}
\end{center}

\vspace{15pt}
\begin{center}
{\large Kaustubh Agashe$\, ^a$, Roberto Contino$\, ^a$,
Alex Pomarol$\, ^b$} \vspace{20pt}

$^{a}$\textit{Department of Physics and Astronomy, Johns Hopkins University \\
 Baltimore, Maryland 21218, USA}

$^{b}$\textit{IFAE, Universitat Aut{\`o}noma de Barcelona, 08193
Bellaterra, Barcelona, Spain}
\end{center}

\vspace{20pt}
\begin{center}
\textbf{Abstract}
\end{center}
\vspace{5pt} {\small \noindent
We study the idea of a composite Higgs in the framework 
of a five-dimensional AdS theory.
We present the minimal model of the
Higgs as a pseudo-Goldstone boson in which electroweak symmetry is broken
dynamically via top loop effects, 
all flavour problems are solved, and contributions to
electroweak precision observables are below
experimental bounds.
Since the 5D theory is weakly coupled,
we are able to fully determine the Higgs potential
and other physical quantities.
The lightest resonances
are expected to have a mass around~$2$~TeV
and should be discovered at the LHC.
The top sector is mostly composite and deviations
from Standard Model couplings are expected.
}

\vfill\eject
\noindent


\section{Introduction and Motivation}

One of the most elegant proposals to explain the origin
of electroweak symmetry breaking (EWSB) is technicolor (TC).
In technicolor theories the breaking of the EW symmetry arises
from a strongly interacting sector,
similarly as the chiral symmetry breaking occurs in QCD.
Simple TC models, however, do not pass the
electroweak precision tests (EWPT) at LEP and SLAC colliders.
The main problem is their contribution to the
Peskin-Takeuchi $S$  parameter \cite{PT} that, when possible to estimate,
is larger than allowed by the experimental data.

An interesting and still economical variation
is to have a Higgs arising as a composite pseudo-Goldstone boson (PGB)
from the strongly interacting sector~\cite{GK}.
In this case, the Higgs mass is protected by an approximate global symmetry
and is only generated via quantum effects.
The electroweak scale is of order $f_\pi$,
where $f_\pi \sim m_\rho/(4\pi)$ is the analog of the pion decay constant and
$m_\rho$ is the scale of the new resonances.
Although this implies a contribution to $S$ similar to TC models,
one needs  to calculate its precise value
to know whether these models pass the EWPT or not.
This has not been possible so far, mainly due to the difficulty of treating
strongly interacting theories. Another problem of these models is 
explaining the origin of fermion masses.

Here we want to pursue the idea of the Higgs as a composite PGB
by studying  it in the framework of five-dimensional models
defined on a slice of AdS spacetime.
Models with an extra dimension have a 4D holographic description
that resembles strongly coupled conformal theories (CFT)
with a large number of ``colors'' $N$.
This description consists in separating the 5D fields in a bulk piece and
a boundary variable, treating them as distinct degrees of freedom.
From the point of view of the 4D effective theory on the boundary,
the bulk acts similarly to a strongly interacting sector,
whose global symmetries are determined by those of the bulk.
Boundary variables are instead equivalent to fields external to the strong sector
and coupled to it. Based on this observation,
one can construct 5D models that mimic
strongly coupled theories.
This correspondence is clearly qualitative in the
sense that we do not know the microscopic dynamics of the strong sector,
but it proves extremely useful to have a quick understanding of the 5D physics
and shape the  low-energy ``chiral'' Lagrangian using symmetry considerations.
Physical quantities can be computed resorting to the full 5D theory, since
it is weakly coupled.
In particular, an expansion in the 5D gauge coupling plays the role of
the $1/N$ expansion of the 4D strongly coupled theory.

In ref.~\cite{Contino:2003ve} a first example of holographic PGB Higgs was given.
Fermion masses were  easily incorporated and a calculation of the Higgs
potential was performed.
Nevertheless, it  lacked a custodial symmetry needed to prevent large
contributions to the Peskin-Takeuchi $T$ parameter.

In this paper we consider an alternative model of Higgs as a
holographic PGB  based on a global SO(5)$\times$U(1)$_{B-L}$ invariance.
This is the minimal symmetry group which contains the 
EW gauge group, can deliver a Higgs as a PGB,
and has an unbroken SO(3) custodial symmetry.
We are able to fully determine the Higgs potential
arising  from one-loop diagrams involving Standard Model (SM) fields
and show  that EWSB is triggered by the top quark contribution.
The Higgs vacuum expectation value (VEV) is given by  $v=\eps f_\pi$, 
where $\eps$ is a model-dependent parameter.
We can accommodate a value of $S$ and $T$ consistent with the EWPT
for $\eps\lesssim 0.4$, which can be obtained
by  a mild tuning in  the parameters of the model  ($\sim 10\%$).
The theory is therefore fully realistic. We find that the
Higgs is always very light, $m_\text{Higgs}\lesssim 140$~GeV
(see fig.~\ref{fig:standard}), one of the most important
predictions of the model. The mild tuning in the parameter space
can be completely eliminated by introducing  an extra
source of SO(5) breaking. In this case we find that the Higgs is
generally heavier (see fig.~\ref{fig:sin4}).
The model also predicts extra gauge and fermionic resonances with
masses $\sim 1-3$ TeV. 
All these new particles come in complete SO(5) multiplets
and have, with the exception of those with top quantum numbers, 
an approximate SO(4) degeneracy.

We proceed as follows.  First (section~\ref{m4dm}), we describe a 4D model
based only on symmetry principles.
The Higgs is assumed to be a composite PGB of a strongly coupled sector
that is conformal at high energies.
By integrating out the CFT dynamics one obtains an effective
Lagrangian for the SM external fields that can be parametrized in
terms of a set of form factors.
The value of these form factors depends on the strong dynamics and
cannot be determined from the 4D theory.
Hence we take a second step (section~\ref{5dmodel}):
we show the corresponding 5D AdS theory
that leads to the same effective Lagrangian
as the  4D model described above.
Since the 5D theory is weakly coupled, we are able to compute
the precise value of the form factors.
This allows us to calculate the $S$ parameter,
the Higgs potential and other important physical quantities.
In section~\ref{sec:EWPT} estimates of the $T$ parameter 
and of the correction to $Z\to b_L\bar
b_L$ are also given using Naive Dimensional Analysis
(NDA), and shown to be close to the experimental limits.
We conclude (section~\ref{conclu})
comparing our model with other popular schemes of EWSB
based on a PGB Higgs.

\section{The minimal 4D model of  PGB Higgs with custodial symmetry}
\label{m4dm}

Let us consider a 4D theory that contains a strongly interacting
sector with  the following properties. It has a large number of
``colors'' $N$,  a mass gap at the infrared scale $\mu_{\rm IR}\sim $TeV, 
and it is conformal at high energies. The mass gap
is responsible for the formation of a tower of bound states with
lowest mass of order $m_\rho\sim\mu_{\rm IR}$. A global symmetry
SU(3)$_c \times$SO(5)$\times$U(1)$_{B-L}$ is spontaneously broken
down to SU(3)$_c \times$SO(4)$\times$U(1)$_{B-L}$ at a scale
$f_\pi \sim (\sqrt{N}/4\pi)\, m_\rho$. 
The operator responsible for this breaking will be
assumed to have a large dimension. The SM gauge bosons and
fermions are elementary fields external to the strongly
interacting CFT. The top quark constitutes an exception and it
will be mostly composite, as we will see later. The SM gauge
bosons couple to the CFT through its conserved currents, gauging
an SU(3)$_c \times$SU(2)$_L\times$U(1)$_Y$ subgroup of the global
invariance. In the following we will neglect the SU(3)$_c$ color
group since it plays no role in the mechanism of EWSB.
Due to the non-linear realization of SO(5), the
theory at tree level has a large set of degenerate vacua, some of
which preserve SU(2)$_L\times$U(1)$_Y$, while others do not. Thus,
whether the electroweak symmetry  is broken or not is a dynamical
issue. By expanding around an SU(2)$_L\times$U(1)$_Y$--preserving
vacuum, so that SO(4)$\sim$SU(2)$_L\times$SU(2)$_R$ and
hypercharge is realized as $Y=T_{3_R}+(B-L)/2$, all fields can be
classified according to their electroweak quantum numbers. In
particular, there is a Goldstone boson transforming as a
\textbf{4} of SO(4), a real bidoublet of SU(2)$_L\times$SU(2)$_R$;
it is a composite state of the CFT and we will identify it with
the Higgs boson.

As long as no explicit breaking of the global SO(5) symmetry
is introduced into the theory, the Higgs field is an exact Goldstone,
and as such it has vanishing potential at any order in perturbation theory.
However, interactions between the CFT and the SM fields
explicitly violate SO(5) and will give rise,
at the one-loop level, to a non-vanishing Higgs potential.
In other words, the degeneracy of classical vacua is lifted by quantum effects
and the Higgs becomes a composite PGB.
The potential generated by gauge interactions alone does not trigger EWSB, 
since gauge forces will tend to align the vacuum along
the SU(2)$_L\times$U(1)$_Y$--preserving direction~\cite{GK}.
A further contribution to the potential, which can ``misalign''
the vacuum, comes however from the fermions living in the elementary sector,
in particular, the top. Fermions will be assumed to couple linearly
to the strong sector through operators ${\cal O}$ made of CFT fields:
${\cal L}=\lambda\, \bar \psi {\cal O}$.
The running coupling $\lambda(\mu)$ obeys the RG equation
\begin{equation}
\mu \frac{d\lambda}{d\mu}=\gamma\, \lambda
 +a\, \frac{N}{16\pi^2}\lambda^3 +\cdots\ ,
\label{rg}
\end{equation}
where the dots stand for terms subleading in the large-$N$ limit,
and $a$ is an ${\cal O}(1)$ positive coefficient. The first term
in eq.~(\ref{rg}) drives the energy scaling for $\lambda$ as
dictated by the anomalous dimension $\gamma=\text{Dim}[{\cal
O}]-5/2$, $\text{Dim}[{\cal O}]$ being the conformal dimension of
the operator ${\cal O}$. The second term originates instead from
the CFT contribution to the fermion wave function renormalization.
The low-energy value of $\lambda$ is determined by $\gamma$. For
$\gamma>0$, the coupling of the elementary fermion to the CFT is
irrelevant, and $\lambda$ decreases with the energy scale $\mu$.
Below $\mu_{\rm IR}$, we have
\begin{equation}
\lambda\sim \left(\frac{\mu_{\rm IR}}{\Lambda} \right)^{\gamma}\,
, \label{la}
\end{equation}
where $\Lambda\sim M_\text{Pl}$ is the UV cutoff of the CFT.
Therefore fermions $\psi$ with
$\gamma>0$  will have a small mixing with the CFT bound-states
and thus a small Yukawa coupling.
For $\gamma<0$, the coupling is relevant and
$\lambda$ flows at low energy towards the fixed-point value
\begin{equation}
\lambda= \frac{4\pi}{\sqrt{N}}\,  \sqrt{\frac{-\gamma}{a}}\, .
\label{lafp}
\end{equation}
In this case the mixing between the fermion $\psi$ and the CFT
is large, and sizable  Yukawa couplings can be generated.

The model is then described by the Lagrangian:
\begin{equation} \label{eq:holoL}
{\cal L} = {\cal L}_\text{CFT} + {\cal L}_\text{SM}
+ J^{a_L\,\mu} W^{a_L}_\mu + J_Y^\mu B_\mu
+ \sum_r \lambda_r\, \bar\psi_r {\cal O}_r + \text{h.c.}\, .
\end{equation}
The sum runs over all SM fermionic representations
$\psi_r=\{ q_L, u_R, d_R, l_L, e_R \}$, (a family index is understood),
and $W_\mu^{a_L}$, ($a_L=1,2,3$), $B_\mu$ stand for SU(2)$_L$ and U(1)$_Y$
gauge bosons respectively.
At tree level the massless spectrum of the theory is that of
the SM.
The Higgs is the Goldstone boson and can be parametrized
 by  the fluctuations along the broken generators
$T^{\hat a}$, $\hat a=1,2,3,4$:
\begin{equation}
\Sigma = \Sigma_0 e^{\Pi/f_\pi} \ ,
 \qquad \Sigma_0 =  (0,0,0,0,1) \ ,
 \qquad \Pi= -i T^{\hat a} h^{\hat a} \sqrt{2} \ .
\end{equation}
Using the SO(5) generators given in appendix~\ref{generators},
one easily finds the explicit expression for $\Sigma$:
\begin{equation} \label{eq:Sigma}
\Sigma = \frac{\sin h/f_\pi}{h} \left( h^1,h^2,h^3,h^4, h \cot h/f_\pi \right) \, ,
 \qquad  h = \sqrt{ (h^{\hat a})^2 } \, .
\end{equation}
The vacuum is characterized by the angular variable
$\langle h\rangle/f_\pi$. Defining  $\eps=\sin\langle h\rangle/f_\pi$, we have
\begin{equation}
\langle \Sigma \rangle = \big(0,0,\eps,0,\sqrt{1-\eps^2}\big)  \, ,
\end{equation}
where the value  of $\eps$ can range between 0 (no EWSB) and 1 (maximal EWSB),
depending on the effective potential of $h$ as we will discuss later.

By integrating out the CFT dynamics, one can write an effective
Lagrangian for the external fields. It is convenient to express
this Lagrangian in an SO(5)-symmetric way.
To do so, we promote the elementary fermions to
fill complete spinorial representations of SO(5).
A spinorial representation of SO(5), a \textbf{4} of SO(5),
contains two (complex) doublets, one transforming under SU(2)$_L$,
the other transforming under SU(2)$_R$. We then embed
$q_L$, $u_R$, $d_R$  as
\begin{equation}
\Psi_{q} = \begin{bmatrix} q_L \\[0.2cm] Q_L \end{bmatrix}\, , \qquad
\Psi_{u} = \begin{bmatrix} q^u_R \\[0.2cm]
             \begin{pmatrix} u_R \\ d'_R \end{pmatrix} \end{bmatrix}\, , \qquad
\Psi_{d} = \begin{bmatrix} q^d_R \\[0.2cm]
             \begin{pmatrix} u'_R \\ d_R \end{pmatrix} \end{bmatrix}\, .
\label{fermio}
\end{equation}
The additional components $Q_L$, $q^u_R$, $d'_R$, $q^d_R$, $u'_R$
must be considered as non-dynamical external sources to the CFT, and they
do not play any physical role.
They are simply a useful tool for embedding the SM fermions in multiplets of SO(5)
and thus being able to write the effective Lagrangian in a compact
SO(5)-invariant fashion.~\footnote{However, choosing a specific representation
for the fermions does have some physical consequence in our theory.
Indeed, we are implicitly demanding for the composite fermionic operators ${\cal O}$
to come in complete multiplets of that particular representation
(a spinorial representation in our specific case).
In our class of theories, these operators have the same quantum numbers
of the CFT fermionic bound states, so that we are implicitly making a specific choice
upon the strong sector.}
Leptons can be promoted to spinorial representations in a similar way, whereas
in the gauge sector we introduce extra non-dynamical vectors to obtain complete
adjoint representations $A_\mu$, $B_\mu$ of SO(5)$\times$U(1)$_{B-L}$.

We can now write the effective Lagrangian using an SO(5)$\times$U(1)$_{B-L}$-invariant
notation. After integrating out all CFT states at tree level,
including fluctuations of the Higgs field around a constant
classical background $\Sigma$, the most general effective
Lagrangian for the external fields is, in momentum space and at
the quadratic level,
\begin{equation}
\begin{split}
{\cal L}_{\rm eff}
 =& \frac{1}{2}P_{\mu\nu}\Big[ \Pi^{B}_0(p)\, B^{\mu} B^{\nu}
    +\Pi_0(p)\, {\rm Tr}\big[A^{\mu} A^{\nu}\big]
    +\Pi_1(p)\, \Sigma A^{\mu} A^{\nu} \Sigma^T\Big] \\
  &+\sum_{r=q,u,d}\bar \Psi_r \pslash \Big[\Pi^{r}_0(p)
 +\Pi_{1}^r(p)\, \Gamma^i\Sigma_i \Big]\Psi_{r}
 +\sum_{r=u,d}\bar \Psi_q \big[M_0^r(p)+M_1^r(p)\, \Gamma^i\Sigma_i\big] \Psi_{r}\, ,
\label{efflag}
\end{split}
\end{equation}
where $P_{\mu\nu}=\eta_{\mu\nu}-p_\mu p_\nu/p^2$ and
 $\Gamma^i$, $i=1,\dots 5$, are the gamma matrices for SO(5)
(see appendix~\ref{generators}):
\begin{equation} \label{matrixS}
\Gamma^i\Sigma_i= 
  \begin{pmatrix}  
    \mathbf{1}\, \cos h/f_\pi & \hat\sigma\, \sin h/f_\pi \\
    \hat\sigma^\dagger\, \sin h/f_\pi & -\mathbf{1}\, \cos h/f_\pi
  \end{pmatrix} \, , \qquad\qquad
  \begin{aligned}
   \hat\sigma &\equiv \sigma^{\hat a}\, h^{\hat a}/h \\
   \sigma^{\hat a} &= \{ \vec \sigma,-i\mathbf{1} \}\, .
 \end{aligned} 
\end{equation}
The form factors $\Pi(p)$, $M(p)$ encode the effect of the strong
dynamics, and cannot be determined perturbatively in the 4D
theory. Their poles match with the CFT spectrum. A possible mixing term
between $\Psi_u$ and $\Psi_d$ in eq.~(\ref{efflag}) has been
neglected since it does not play any role in our calculations.
Also, we have not written down possible bare kinetic terms and
gauge-fixing terms for the external fields, i.e. terms not induced
by the strong dynamics. They can be included in a straightforward
way. We are only interested in two-point functions since, as we
will see below, these are the only ones needed for the calculation
of the $S$ parameter and the Higgs potential.

From the form factors of eq.~(\ref{efflag}) one can derive the
low-energy effective theory. This is the theory of the light
states, the SM fields and the Higgs (the equivalent of the chiral
theory in QCD). It is obtained by performing  an expansion in
derivatives and light fields over $m_\rho$:
\begin{equation}
{\cal L} ={\cal L}_\text{kin} +{\cal L}_\text{yuk}
 -V(\Sigma) + \Delta{\cal L} \, .
\label{eq:effL}
\end{equation}
The term  ${\cal L}_\text{kin}$ contains the kinetic terms of the
dynamical fields
\begin{equation}
{\cal L}_\text{kin}=
 \frac{f^2_\pi}{2} \left(D_\mu \Sigma\right) \left(D^\mu \Sigma\right)^T
 + \sum_r Z_r\, \bar\psi_r \not\!\! D
\psi_r - \frac{1}{4g^2} W^{a_L}_{\mu\nu} W^{a_L\, \mu\nu}
 - \frac{1}{4g^{\prime\, 2}} B_{\mu\nu} B^{\mu\nu}\, ,
\end{equation}
where $Z_q=\Pi_0^q(0)+\Pi_1^q(0)$, $Z_{u,d}=\Pi_0^{u,d}(0)-\Pi_1^{u,d}(0)$,
$f_\pi^2=\Pi_1(0)$, $1/g^2=-\Pi'_0(0)$, 
$1/g^{\prime\, 2}=-( \Pi^{B\,\prime}_0(0) + \Pi'_0(0) )$.
The Higgs potential $V(\Sigma)$ is generated at one loop by gauge
and fermion interactions.
We will show below that the top contribution can trigger EWSB
and the Higgs field $h$ acquires a VEV, breaking SO(4) down to
the custodial SO(3) group.
From the kinetic term for $\Sigma$ we obtain
$M_W^2 = g^2 v^2/4$, where we have defined the EWSB scale
\begin{equation} \label{eq:v}
v \equiv \epsilon f_\pi
= f_\pi \sin \frac{\langle h\rangle}{f_\pi}=246\ {\rm GeV} \, .
\end{equation}
The term ${\cal L}_\text{yuk}$ contains the
Yukawa couplings between the Higgs and the elementary fermions and comes
from the expansion of the last term  of eq.~(\ref{efflag}):
\begin{equation}
{\cal L}_\text{yuk}
= \frac{\sin h/f_\pi}{h} \left[ M_1^u(0)\,
 \bar q_L h^{\hat a} \sigma^{\hat a}
   \begin{pmatrix} u_R \\ 0 \end{pmatrix} + M_1^d(0)\, \bar q_L h^{\hat a} \sigma^{\hat a}
   \begin{pmatrix} 0 \\ d_R \end{pmatrix}+ \text{h.c.} \right] \, .
\end{equation}
When the Higgs acquires a VEV, the fermions get a mass
\begin{equation}
m_{u,d}= \frac{M_1^{u,d}(0)}{\sqrt{Z_q Z_{u,d}}}\, \frac{v}{f_\pi} \equiv y_{u,d}\, v\, ,
\end{equation}
where by NDA $y_{u,d}\sim \lambda_{u,d} \lambda_q\, \sqrt{N}/4\pi$.
By choosing $\gamma_{q,u,d}>0$, we have, according to eq.~(\ref{la}),
that $\lambda_{q,u,d}$ are strongly suppressed at low energies,
and the fermions are weakly coupled to the CFT.
This can be used to explain in a natural way the smallness and the hierarchical
structure of the masses of the light
fermions~\cite{Grossman:1999ra,Gherghetta:2000qt}:
\begin{equation}
m_{u,d}\sim \frac{\sqrt{N}}{4\pi}\left(\frac{\mu_{\rm
IR}}{\Lambda} \right)^{\gamma_q+\gamma_{u,d}}v\, . \label{lightf}
\end{equation}
It is interesting to notice that this theory has a GIM mechanism,
since
flavour changing neutral current (FCNC) effects
involving light fermions are also suppressed by the couplings $\lambda_{u,d,q}$
(see, for example~\cite{Gherghetta:2000qt,Huber:2000ie,Agashe:2003zs,Agashe:2004cp}).
In order to have a large top mass, we will require
$\gamma_u<0$ and $\gamma_q\simeq 0$ for the third quark generation.
This choice is compatible with EWPT, as we will discuss in detail, and
implies that the physical right-handed top quark is mostly composite.

The last term $\Delta {\cal L}$ in the effective Lagrangian
(\ref{eq:effL}) contains all higher-order operators
in the chiral expansion.
The only one that is relevant for us here is
that responsible for the $S$ parameter. It originates
from the third term of eq.~(\ref{efflag}):
\begin{equation}
\Delta {\cal L} \supset  \frac{1}{2}\Pi^\prime_1(0) \,
W_{\mu\nu}^{a_L} B^{\mu\nu}\, \Sigma T^{a_L} Y\Sigma^T\, ,
 \label{delta}
\end{equation}
where $T^{a_L}$, $Y$ are respectively the generators of SU(2)$_L$
and hypercharge.
Eq.~(\ref{delta}) gives
\begin{equation}
S =4\pi \Pi^\prime_{1}(0)\eps^2 \, .
\label{spa}
\end{equation}
The $T$ parameter does not receive any contribution at tree level
from the CFT due to the custodial symmetry.
Nevertheless, it  can be induced at the quantum
level due to top interactions.
We will discuss
in section~\ref{sec:EWPT} the size of these contributions.
Apart from $S$ and $T$, there are other two parameters
constrained by LEP: $W$ and $Y$, defined in \cite{Barbieri:2004qk}.
They are however quite small in the present model, since
they arise from dimension-six operators and are thus suppressed by a factor
$(g^2 f_\pi^2/m^2_\rho)$ compared to $S$ and $T$.

\subsection{Higgs potential and vacuum misalignment}
\label{vacmis}

A virtual exchange of elementary fields can transmit the explicit
breaking of SO(5) from the elementary sector to the CFT and
generate a potential for the PGB Higgs. The dominant
contribution comes at one-loop level from the elementary SU(2)$_L$ gauge
bosons and top quark.
This is given by
the  Coleman-Weinberg potential
\begin{equation}
V(h) =  \frac{9}{2} \int\! \frac{d^4p}{(2\pi)^4}\, \log\Pi_W
 - (2  N_c) \int\! \frac{d^4p}{(2\pi)^4}\,
 \bigg[ \log\Pi_{b_L} + \log\left(p^2 \Pi_{t_L}\Pi_{t_R}-\Pi_{t_Lt_R}^2\right)\bigg]\, ,
\label{pot}
\end{equation}
where $\Pi_i(p)$ are the self-energies of the corresponding SM fields
in the background of $h$. These can
can be written as functions of the form factors of
eq.~(\ref{efflag}), by using eq.~(\ref{matrixS}):
\begin{equation} \label{fff}
\begin{split}
\Pi_W &= \Pi_0+ \frac{\Pi_1}{4} \sin^2\frac{h}{f_\pi}\, , \\
\Pi_{t_Lt_R} &= M_1^u\sin\frac{h}{f_\pi}\, ,
\end{split} \qquad \quad
\begin{split}
\Pi_{b_L} &= \Pi_{t_L}= \Pi_0^q+\Pi_1^q\cos\frac{h}{f_\pi}\, , \\
\Pi_{t_R} &= \Pi_0^u-\Pi_1^u\cos\frac{h}{f_\pi} \, .
\end{split}
\end{equation}
Apart from a constant piece, the  potential of eq.~(\ref{pot}) is
finite since the form factors $\Pi_1$ and $M_1$  drop with the
momentum as $|\langle\Phi\rangle|^2/p^{2d}$, where $\Phi$ is the
CFT operator of dimension $d\gg 1$ responsible for the SO(5)
breaking.~\footnote{In fact, in the 5D model the form factors drop
exponentially with the momentum, corresponding to
$d\rightarrow\infty$.} This fast decrease with the momentum allows
us to expand the logarithms in eq.~(\ref{pot}) and write the
approximate formula~\footnote{This approximate formula leaves out
 the  top logarithmic contribution to the Higgs quartic
coupling $\propto \log(m_t/m_\rho) \sim \log \eps$ since it comes
from a subleading term in the expansion. This
contribution can be large if  $\eps$ is very small, and in that
case it should be incorporated. For the qualitative discussion
presented here, we will neglect it. For
the 5D calculation of the next section, however,
we will take the full potential eq.~(\ref{pot}).}
\begin{equation} \label{apxpot}
V(h)\simeq \alpha\cos\frac{h}{f_\pi}-\beta\sin^2\frac{h}{f_\pi}\, ,
\end{equation}
where 
\begin{equation} \label{ab}
\alpha = 2N_c \int\! \frac{d^4p}{(2\pi)^4}\, 
 \left( \frac{\Pi^u_1}{\Pi^u_0} -2 \frac{\Pi^q_1}{\Pi^q_0} \right)\, , \qquad
\beta  = \int\! \frac{d^4p}{(2\pi)^4}\,
 \left( 2N_c \frac{(M_1^u)^2}{(-p^2)\,\Pi^q_0\Pi^u_0} 
        - \frac{9}{8}\frac{\Pi_1}{\Pi_0} \right)\, .
\end{equation}
This potential has a minimum at $\cos
h/f_\pi=-\alpha/(2\beta)$, i.e.
\begin{equation}
\epsilon= \sqrt{1-\left(\frac{\alpha}{2\beta}\right)^2}\, .
\label{minimum}
\end{equation}
Thus, for suitable values of $\alpha$ and $\beta$ the EWSB can
occur dynamically. Gauge fields only contribute to the $\sin^2$
operator with an overall positive coefficient, $\beta_\text{gauge}
< 0$, and tend to align the vacuum in the SU(2)$_L$ preserving
direction. A misalignment of the vacuum can only come from top
loops, and only if the coefficients $\alpha$ and $\beta$ are
comparable in size. An NDA estimate of the two coefficients shows
that $\alpha$ is expected to be larger than $\beta$: $\alpha \sim
m_\rho^4\, \lambda^2_{q,u}\, N_c\, N/(16\pi^2)^2$, $\beta\sim
m^4_\rho\, y_t^2\,N_c\, N/(16\pi^2)^2$, where $y_t\sim \lambda_q
\lambda_u \sqrt{N}/4\pi$ is the top Yukawa coupling. However,
$\alpha$ turns out to be generally smaller than its NDA estimate,
and the contributions from $q_L$ and $t_R$ are always opposite
in sign (see eq.~(\ref{fff}) and~(\ref{ab})), thus tending naturally 
to cancel each other.
The physical Higgs mass is given~by
\begin{equation} \label{mHiggs}
m_{\rm Higgs}^2 \simeq\frac{2\beta\, \epsilon^2}{f^2_\pi}
\sim \frac{2N_c}{N}\, y_t^2\, v^2\, .
\end{equation}
We see that for moderate values of $N$ the Higgs mass can be
above the experimental bound $m_{\rm Higgs}>114$~GeV.
Notice that, despite being generated at one-loop level,
the quartic coupling is ${\cal O}(1)$.
Indeed, due to the strong dynamics involved in the loop
diagrams, the actual loop expansion parameter 
is $1/N$, as it is evident from eq.~(\ref{mHiggs}).

Remarkably, this minimal theory can accomplish a realistic EWSB.
To quantify this statement we must calculate the precise value
of $\alpha$, $\beta$  and the $S$ parameter.
This will be done in the 5D theory.
We will see
that in order to satisfy the experimental constraints coming from $S$, $T$ and
$Z\to b\bar b$, a value $\eps\lesssim 0.4$ is necessary.
This implies, from eq.~(\ref{minimum}), that the relation $\alpha\simeq 2\beta$
must be fulfilled at the  $10\%$ level.
We will show that this can be  accomplished
in certain regions of the 5D parameter space.

It is interesting to notice that the mild tuning required in the
minimal model can be completely avoided if one introduces new
sources of SO(5) breaking that generate extra terms in the Higgs
potential. For example, if the Higgs potential has an extra
$\sin^4$ term
\begin{equation} \label{sin4-sin2pot}
V(h)\simeq \alpha\cos\frac{h}{f_\pi}-\beta\sin^2\frac{h}{f_\pi}+
\gamma\, \sin^4 \frac{h}{f_\pi}\, ,
\end{equation}
the minimum, for $\eps<1$, is given by
\begin{equation} \label{newminimum}
\eps^2 = \frac{\alpha}{4\gamma}\, \left[
 \frac{\pm 1 + 2\beta/\alpha}{1+\beta/4\gamma} \right] + {\cal O}(\eps^4) \, ,
\end{equation}
where the difference in sign comes from  $\cos h/f_\pi \simeq \pm
(1 -\eps^2/2)$. Eq.~(\ref{newminimum}) shows that a coefficient
$\gamma$ slightly larger than $\alpha$, but still of one loop
size, can give $\eps\sim 0.2-0.4$ with virtually no fine tuning in
the second factor. New contributions to the potential can come
from different sources. In appendix~\ref{tree} we give an  example of how
eq.~(\ref{sin4-sin2pot}) can be generated by loop effects
of additional fermions, or at tree-level by
an extra scalar field, following a mechanism already proposed in
ref.~\cite{Contino:2003ve}.

Comparing with the original Georgi-Kaplan models
where a large hierarchy $v\ll f_\pi$ (and therefore a large fine-tuning)
was required to avoid the strong constraints from FCNC~\cite{GK},
we see that in our case we only need a very small hierarchy between
$v$ and $f_\pi$. No relevant constraint on $\eps$ comes in our model
from FCNC due to the different realization of flavour.

\section{The  5D model}
\label{5dmodel}

The 4D theory presented above can be obtained as the holographic description
of a 5D weakly coupled model.
In this section we describe the 5D model and show how to compute the
form factors of eq.~(\ref{efflag}).

The 5D spacetime metric is given by~\cite{Randall:1999ee}
\begin{equation}
  ds^2 = \frac{1}{(kz)^2} \left(\eta_{\mu\nu}\, dx^\mu dx^\nu - dz^2\right)
  \equiv g_{MN}\, dx^M dx^N\, ,
\label{eq:metric}
\end{equation}
where the 5D coordinates are labeled by capital Latin letters,
$M=(\mu,5)$, with $\mu=0,\dots,3$; $z = x^5$ represents the
coordinate for the fifth dimension and $1/k$ is the AdS curvature
radius. This spacetime has two boundaries at $z=L_0 \equiv 1/k
\sim 1/M_{\rm Pl}$ (UV-brane) and at $z = L_1 \sim \mu_{\rm IR}
\sim 1/{\rm TeV}$ (IR-brane). The theory is defined on the line
segment $L_0 \leq z \leq L_1$. The gauge symmetry in the 5D bulk
is taken to be SU(3)$_c \times$SO(5)$\times$U(1)$_{B-L}$ reduced
to SU(3)$_c \times$SO(4)$\times$U(1)$_{B-L}$ on the IR-brane and
to SU(3)$_c \times$SU(2)$_L\times$U(1)$_Y$ on the UV-brane.  This
can be easily achieved by imposing a Dirichlet boundary condition
for the gauge bosons whose generators we want to break. We choose
to work in the unitary gauge $\partial_z (A_5/z)=0$ (see
ref.~\cite{Contino:2003ve}), where only physical gauge
configurations survive. In this gauge $A_5$ is non-vanishing only
in its SO(5)/SO(4) components, which are however constrained to
have a fixed profile along the fifth dimension: $A_5(x,z)=
\zeta(z) h(x)$, $\zeta(z)= z\sqrt{2/(L_1^2-L_0^2)}$. Thus,
physical fluctuations of $A_5$ correspond to a 4D scalar field
$h(x)$ transforming as a {\bf 4} of SO(4), the Higgs. From the
point of view of the 5D theory, a potential for $A_5$ is forbidden
at tree level by gauge invariance, but it is generated radiatively
as a finite-volume effect from non-local operators. This is the
Hosotani mechanism for symmetry breaking \cite{hosotani}.

The SM fermions are embedded into 5D Dirac spinors $\xi_i$ that live in the bulk
and belong to the ${\bf 4_{1/3}}$ representation of SO(5)$\times$U(1)$_{B-L}$.
For each quark family we define
\begin{equation} \label{fstates}
\begin{split}
\xi_q &= \begin{bmatrix}
  q_L (++) & q_R (--)\\[0.1cm] Q_L(--) & Q_R (++) \end{bmatrix}\, , \quad
\xi_u = \begin{bmatrix}
  q_L^u (+-)  & q_{R}^{u}(-+) \\[0.1cm]
  Q_L^u = \begin{bmatrix} u^{c\, \prime}_L (-+) \\ d^{c\, \prime}_L (++)\end{bmatrix}
 &Q^u_R =\begin{bmatrix} u_R (+-) \\ d^{\,\prime}_R (--)\end{bmatrix}
        \end{bmatrix}\, , \\
\xi_d &= \begin{bmatrix}
  q_L^d (+-) & q_{R}^{d}(-+) \\[0.1cm]
  Q_L^d=\begin{bmatrix} u^{c\,\prime \prime}_L(++) \\ d^{c\, \prime \prime}_L (-+)\end{bmatrix}
 &Q^d_R =\begin{bmatrix} u^\prime_R (--) \\ d_R (+-)\end{bmatrix}
        \end{bmatrix}\, ,
\end{split}
\end{equation}
where leptons are realized in a similar way.
Here $(\pm,\pm)$ is a shorthand notation to denote a Neumann ($+$)
or Dirichlet ($-$)
boundary condition on each brane.
Chiralities under the 4D Lorentz group have been denoted with $L,R$, while
small $q$'s (capital $Q$'s) denote doublets under SU(2)$_L$ (SU(2)$_R$).
Massless modes in eq.~(\ref{fstates}) arise from $(+,+)$ fields. These are
$q_L$, $Q_R$ and $d^{c\,\prime}_L$, $u^{c\,\prime \prime}_L$.
We get rid of the latter two states by adding an extra field on the IR-brane,
$\widetilde Q_R$, which marries them:
$[ \bar Q_L^u+ \bar Q_L^d]\widetilde Q_R$.

Gauge invariance, however, allows one to write mass
and kinetic mixing terms among the different
multiplets $\xi_q$, $\xi_u$, $\xi_d$ in the bulk and on the IR-brane.
The bulk kinetic and mass matrices are SO(5)--symmetric and can be
always simultaneously
diagonalized through a field redefinition. In that basis, the most general
SO(4)-invariant set of mass terms one can write on the IR-brane includes
\begin{equation}
\label{massmixing}
[ \widetilde M_u
\bar Q_L^u +  \widetilde M_d
 \bar Q_L^d] Q_R +
\bar q_L[ \widetilde m_u
 q_R^u + \widetilde m_d  q_R^d] + h.c.
\end{equation}
Therefore, the massless states
 become a mixture of $q_L$ with $q_L^u$ and $q_L^d$;
$u_R$ with the upper component of $Q_R$; $d_R$ with the lower component of $Q_R$.
The mixing angle depends on the value of the 5D bulk masses $M_5^i=c_i k$.
Yukawa couplings to $A_5$ arise from
the 5D covariant derivative in the fermion bulk
kinetic terms.  This means that $A_5$
can only connect SU(2)$_L$ doublets with SU(2)$_R$ doublets of opposite Lorentz chirality
inside the same 5D multiplet (for example $q_L$ with $Q_R$).
Yukawas for the physical
massless modes then proceed through the
mixing and can be suppressed depending on the
mixing angle.

\subsection{The holographic approach: Matching to the 4D theory}
\label{sec:effL}

The 5D model described above
has  exactly the same properties as the 4D CFT theory
described in section~\ref{m4dm}.
In particular it leads to the same effective Lagrangian
eq.~(\ref{efflag}) and to the same
low-energy theory eq.~(\ref{eq:effL}).
In order to match the two theories,
it is more convenient to follow the holographic approach
instead of the more popular Kaluza-Klein (KK) decomposition.

The holographic procedure consists in separating the bulk fields
from their UV-brane value and treating them as distinct variables.
If we integrate out the bulk with fixed values of the fields on
the UV-brane,  we obtain a 4D (non-local) theory defined on the UV
boundary. This is the theory to be matched with
eq.~(\ref{efflag}). Boundary values of 5D fields 
with Neumann (Dirichlet) condition on the UV-brane correspond to the
external dynamical (non-dynamical) fields of eq.~(\ref{efflag}), 
while the bulk plays the role of the CFT sector. 
The holographic procedure is clearly
inspired by the AdS/CFT correspondence~\cite{Maldacena:1997re},
although it does not rely in any sense on the validity of that
conjecture. It simply represents an alternative way of describing
the 5D model, motivated by symmetry principles. The symmetries of
the bulk are the same as those of the CFT, while the 4D UV-brane
sector respects only the SM gauge invariance (see
refs.~\cite{Barbieri:2003pr} and~\cite{Contino:2004vy} for a
similar application of the holographic description).

To obtain the 5D prediction for the form factors of
eq.~(\ref{efflag}) we match the two theories on the
SO(4)-invariant vacuum: $\Sigma=\Sigma_0$ (i.e. $h=0$). Let us
start with the gauge sector. On the AdS$_5$ side, after
integrating the bulk as a function of the UV-boundary fields, one
finds (at the quadratic level),
\begin{equation}
{\cal L}_{\rm eff}=\frac{1}{2}P_{\mu\nu}\big[\Pi_{a}(p)
 A^{a\, \mu}   A^{a\, \nu} + 
\Pi_{\hat a}(p) A^{\hat a\, \mu}A^{\hat a\, \nu}\big]\, .
\label{efflads}
\end{equation}
The indexes $a$ and $\hat a$ run respectively  over the SO(4)
generators (unbroken on the IR-brane) and the SO(5)$/$SO(4)
generators (broken on the IR-brane), and
\begin{equation} \label{sigma}
\Pi_{a,\hat a}(p)= - \frac{1}{g^2_5 k} \frac{p}{L_0}\,
 \frac{Y_{0}(pL_0) \tilde J_{0,1}(pL_1)- \tilde Y_{0,1}(pL_1) J_{0}(pL_0)}
      {Y_{1}(pL_0) \tilde J_{0,1}(pL_1)- \tilde Y_{0,1}(pL_1) J_{1}(pL_0)}\, ,
\end{equation}
with $\tilde J_{1}(pL_1)=J_{1}(pL_1)$, $\tilde
J_{0}(pL_1)=J_{0}(pL_1) - (g_5^2 k/g_{\rm IR}^2) \,pL_1\,
J_{1}(pL_1)$ and similarly for $\tilde Y_{0,1}$. Here $g^2_5$
denotes the SO(5) bulk gauge coupling, and $1/g^2_{\rm IR}$ is the
coefficient of the SO(4) boundary kinetic term on the IR-brane. 
Analogous formulas apply for the U(1)$_{B-L}$ gauge boson,
that has been omitted in eq.~(\ref{efflads}) for simplicity. 
We have not written down possible
boundary kinetic terms on the UV-brane, though they can be
included in a straightforward way. Eq.~(\ref{efflads}) must be
matched to eq.~(\ref{efflag}) after setting
$\Sigma\rightarrow\Sigma_0$. As we said, boundary values of 5D
fields with Neumann (Dirichlet) conditions on the UV-brane must be
identified with the dynamical (non-dynamical) external fields of
the 4D theory. We get
\begin{equation}
\Pi_a(p)=\Pi_0(p)\, ,\qquad   \Pi_{\hat a}(p)=\Pi_0(p)+\frac{1}{2}\Pi_1(p)\, ,
\end{equation}
that leads to the determination of the gauge form factors:
\begin{equation}
\Pi_0(p)=\Pi_a(p)\ , \qquad
\Pi_1(p)=2\big[\Pi_{\hat a}(p)-\Pi_a(p)\big]\, .
\label{ff}
\end{equation}
For the fermions we can proceed in a similar way.
We integrate out the bulk fermionic fields as a function
of their values on the UV-brane.
We are only allowed to  fix on the boundary either the left- or the right-handed
fermionic component (left- or right-handed source description, see~\cite{Contino:2004vy}).
For $\xi_q$ we will take the left-handed component,
$\xi_{q\, L}=(q_L,Q_L)^T$, while for $\xi_u$ we will take the right-handed
component, $\xi_{u\, R}=(q^u_R,Q^u_R = [u_R,d^\prime_R])^T$.
We omit $\xi_d$ for simplicity.
At the quadratic level we obtain
\begin{equation} \label{selfermions}
\begin{split}
{\cal L}_{\rm eff} =&
   \Pi_{q_L}(p)\, \bar q_L \pslash  q_L + \Pi_{Q_L}(p)\, \bar Q_L \pslash  Q_L
  +\Pi_{q^u_R}(p)\, \bar q^u_R \pslash  q^u_R + \Pi_{Q^u_R}(p)\, \bar Q^u_R \pslash  Q^u_R \\
 &+ M_{q}(p)\, \bar q_L q^u_R + M_{Q}(p)\, \bar Q_L Q^u_R\, .
\end{split}
\end{equation}
The fermionic self-energies $\Pi_i(p)$ and $M_i(p)$
are given in appendix~\ref{fselfen} in terms of brane-to-brane 5D propagators.
Matching with the 4D CFT theory, we have
\begin{equation} \label{selff}
\begin{split}
\Pi_{0,1}^q(p)   &= \frac{1}{2} \left[ \Pi_{q_L}(p)\pm\Pi_{Q_L}(p)\right] \, ,  \\
\Pi_{0,1}^{u}(p) &= \frac{1}{2} \left[ \Pi_{q^u_R}(p)\pm\Pi_{Q^u_R}(p)\right] \, , \\
M_{0,1}^u(p)     &= \frac{1}{2} \left[M_{q}(p) \pm M_{Q}(p)\right] \, .
\end{split}
\end{equation}
The anomalous dimensions $\gamma$ of the 4D CFT theory
are related to the 5D fermion masses $M_5^i=c_i k$ according
to~\footnote{The different relation for $q$ and $u,d$ is because
we are fixing on the UV-boundary  the left-handed component for $q$
and the right-handed  component for $u$,$d$.
See ref.~\cite{Contino:2004vy} for details.}
\begin{equation}
\label{gac} \gamma_q=\left|c_q+\frac{1}{2}\right|-1\ ,
\qquad \gamma_{u,d}=\left|c_{u,d}-\frac{1}{2}\right|-1\,
.
\end{equation}
Therefore the requirement $\gamma_{q,u,d}>0$ for the light fermions
(see above eq.~(\ref{lightf})) implies $c_q>1/2$ and $c_{u,d}<-1/2$,
while for the top $\gamma_{q}\simeq 0$ and $\gamma_{u}<0$
implies $c_q\simeq 1/2$ and  $|c_u|< 1/2$.

It is useful at this point to define
the number of CFT colors $N$ by identifying $1/N$
with the perturbative expansion parameter in 5D:
\begin{equation}
\frac{1}{N} \equiv \frac{g_5^2 k}{16\pi^2}\, .
\label{n}
\end{equation}
The  5D NDA condition for calculability reads
\begin{equation}
\frac{\pi k}{\Lambda_S} =\frac{g_5^2 k}{24\pi^2}  \ll 1\, ,
\end{equation}
where $\Lambda_S$ is the strong cutoff scale of the 5D theory.
Using the relation $1/g^2=\ln(L_1/L_0)/g_5^2k+1/g^2_{\rm UV}+1/g^2_{\rm IR}$,
where $1/g_{\rm UV}$ ($1/g_{\rm IR}$) is the SU(2)$_L$ UV-brane
(SO(4) IR-brane) kinetic term and $g\simeq 0.65$ is the effective
SU(2)$_L$ coupling,
we can derive a rough lower bound for $g^2_5k$:
\begin{equation} \label{g5large}
{g_5^2 k} > g^2 \ln(L_1/L_0) \sim 16\, .
\end{equation}
Therefore $N$ is restricted in the interval $1\ll N\lesssim 10$.

\subsection{Predictions of the 5D model}
\label{sec:analysis}

Having matched the 5D and 4D theories, we can now derive
the predictions of  the 5D model.
First, we can obtain from eq.~(\ref{sigma})
the mass spectrum and decay constants of the vector resonances
of the theory. For this purpose we decompose $\Pi_{a,\hat a}(p)$ as
an infinite sum over narrow mesons, as prescribed by the large-$N$ limit:
\begin{equation}
\Pi_{a}(p)=  p^2
 \sum_n\frac{F^2_{\rho_n}}{p^2+m^2_{\rho_n}}\, ,\qquad
\Pi_{\hat a}(p)= p^2
 \sum_n\frac{F^2_{a_n}}{p^2+m^2_{a_n}} + \frac{1}{2} f^2_\pi\, .
\label{valargen}
\end{equation}
Eq.~(\ref{sigma}) thus gives
\begin{equation} \label{eq:scales}
f_\pi^2=\frac{4}{g_5^2 k}\, \frac{1}{L^2_1}\, , \qquad
m_\rho\equiv m_{\rho_1}\simeq
  \frac{3\pi/4}{\sqrt{1+9\pi^2/32\, z_{\rm IR}}}\, \frac{1}{L_1}\, ,
\qquad  m_{a_1}\simeq \frac{5\pi}{4}\frac{1}{L_1}\, ,
\end{equation}
where we have defined $z_{\rm IR}=g^2_5k/g^2_{\rm IR}$.
From  eq.~(\ref{spa}) and (\ref{ff}) we can obtain the prediction for the $S$
parameter:
\begin{equation}
S = \frac{3}{8} \frac{N}{\pi} \epsilon^2
\left[1+\frac{4}{3}\, z_{\rm IR} \right]\, .
\end{equation}
The $99\%$ CL experimental bound from LEP,
$S\lesssim 0.3$~\cite{Barbieri:2004qk}~\footnote{It corresponds to
an extra contribution to the $\eps_3$ parameter~\cite{eps123}
$\Delta\eps_3\lesssim 2.5 \cdot 10^{-3}$
relative to the SM value with $m_\text{Higgs}$=115~GeV.}, requires:
\begin{equation}
\epsilon\,\lesssim 0.5\,
\sqrt{\left(\frac{10}{N}\right)\frac{1}{1+4/3\, z_{\rm IR}}}\, .
\label{cons}
\end{equation}
We see that in order to remain in the weak coupling regime, $1/N\ll 1$,
we need $\epsilon< 1$.
In particular, taking $N=10$ and  $z_{\rm IR}=0\ (1)$ requires
$\epsilon\sim 0.5\ (0.3)$.
A way of recasting the result for $S$ is by fixing
$f_\pi\eps =v$ according to eq.~(\ref{eq:v}),
and using the value of $m_\rho$ as computed in eq.~(\ref{eq:scales}).
One obtains
\begin{equation}
\label{rhos}
 S \simeq \frac{27\pi^3}{32}
\left(\frac{1+4/3\, z_{\rm IR}}{1+9\pi^2/32\, z_{\rm IR}}\right)
\frac{v^2}{m_\rho^2}
\, .
\end{equation}
In order to satisfy  $S\lesssim 0.3$, the lowest vector state cannot be  lighter
than $m_\rho\sim 2.3\ (1.6)$ TeV for $z_{\rm IR}=0\ (\infty)$.

\begin{figure}
\begin{minipage}[t]{0.95\linewidth}
\epsfig{file=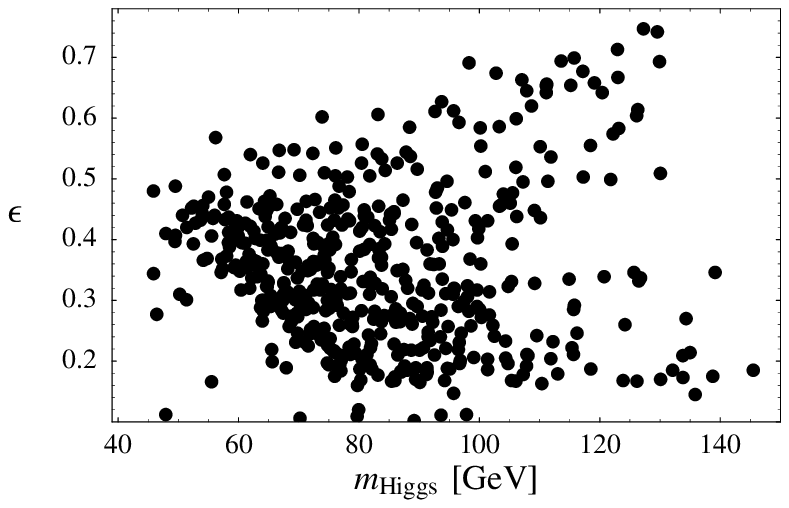,width=0.49\linewidth} \qquad
\epsfig{file=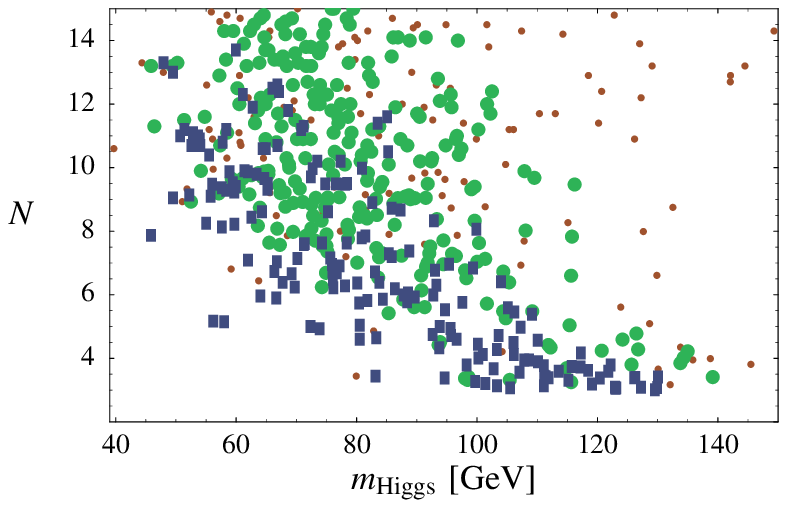,width=0.49\linewidth}
\end{minipage}
\caption{\it Scatter plots in the $(m_\text{Higgs},\eps)$ plane (left) and in the 
$(m_\text{Higgs},N)$ plane (right), obtained by scanning over the input parameters
in the minimal model.
In the second plot, blue squares correspond to $\eps>0.4$, 
green fat dots to $0.2<\eps<0.4$, small red dots to $\eps<0.2$.}
\label{fig:standard}
\end{figure}

To determine whether EWSB is triggered and an $\eps< 1$ is
generated in our model, one has to compute and minimize the Higgs
potential (\ref{pot}). The potential is completely determined by
the self-energies eqs.~(\ref{sigma}) and
(\ref{FFfirst})--(\ref{FFlast}), which are in turn functions of
few input parameters: the bulk SO(5) gauge coupling $g_5$,  the
gauge kinetic terms on the UV and IR branes, $1/g^2_\text{UV}$ and
$1/g^2_\text{IR}$ (respectively for SU(2)$_L$ and SO(4)), 
and the top bulk and IR-brane masses, $c_q$, $c_u$, $\m$ and $\M$.
 For simplicity we have neglected the subleading contribution to
the potential coming from the bottom quark and the hypercharge
gauge field. Performing a numerical analysis, we found that in a
large region of the input parameter space EWSB is indeed triggered
and the bound (\ref{cons}) on the $S$ parameter is satisfied. The
results are summarized in fig.~\ref{fig:standard}, where $\eps$
and the number of colors of the CFT  defined by (\ref{n}), 
are given as a function of the physical Higgs
mass. Only points which satisfy the bound $S< 0.3$ are shown. The
values of the input parameters have been chosen as follows. We
fixed the UV coupling $g_\text{UV}$ by requiring that the
low-energy SU(2)$_L$ gauge coupling $g$ equals its experimental
value, while the IR gauge kinetic term has been set to be of loop
order, $1/g^2_\text{IR}=1/16\pi^2$. The value of $N$ is extracted
by fixing the top mass to its experimental value $m_t^\text{exp} =
178.0 \pm 4.3$ GeV~\cite{Azzi:2004rc}. The top bulk masses $c_q$,
$c_u$ must lie in the interval $-1/2 < c_q,c_u < +1/2$, otherwise
the coupling between the elementary $q_L$, $t_R$ and the CFT
becomes irrelevant, implying a too small top mass. Values of $c_q$
close to $+1/2$ seem to be preferred in order not to have large
corrections to $Z\to b\bar b$, as we will explain in detail in the
next section. Also, a $c_u$ too close to $+1/2$ is disfavored by
$Z\to b\bar b$, while values of $c_u$ close to $-1/2$ seem to give
a smaller $T$ parameter and they are preferred. We thus scan in
our numerical analysis over $-0.3 < c_u < +0.3$ and  $0.3 < c_q < 0.5$.
Finally, for the IR-brane masses we have taken the following values: 
$ \m= 0.5, 1.0, 1.5$, and $-2.1 < \M < -0.4$.

\begin{figure}[t]
\begin{minipage}[t]{0.95\linewidth}
 \epsfig{file=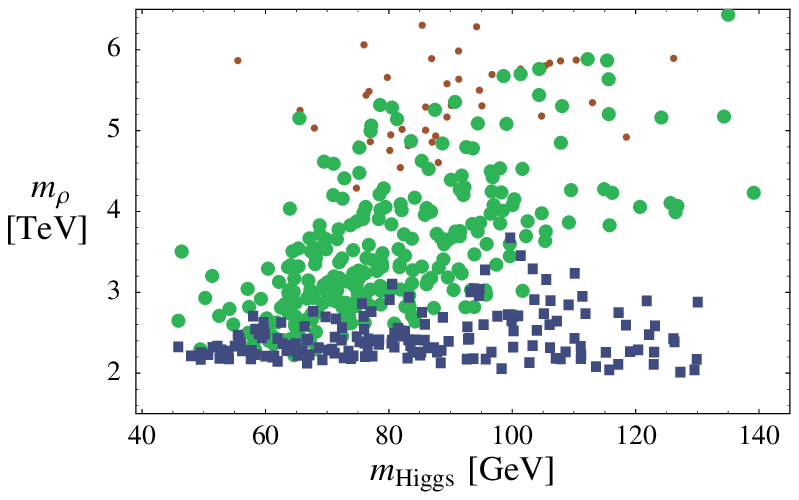,width=0.49\linewidth} \qquad
\epsfig{file=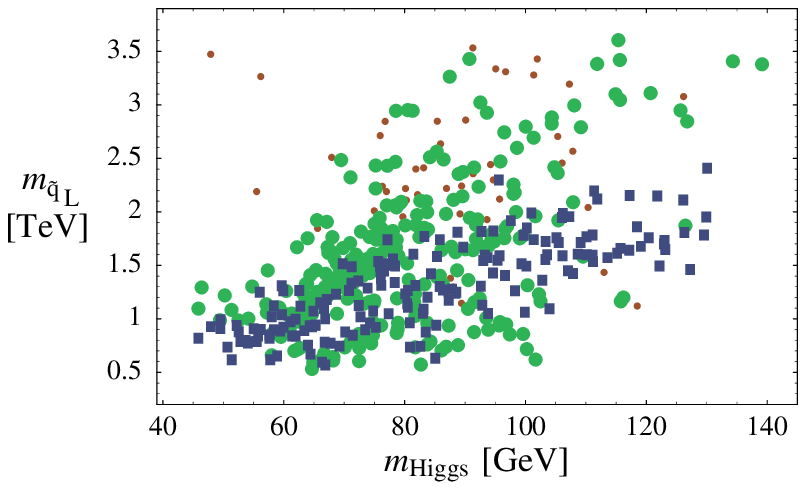,width=0.49\linewidth}
\end{minipage} \\[0.2cm]
\begin{minipage}[t]{0.95\linewidth}
\epsfig{file=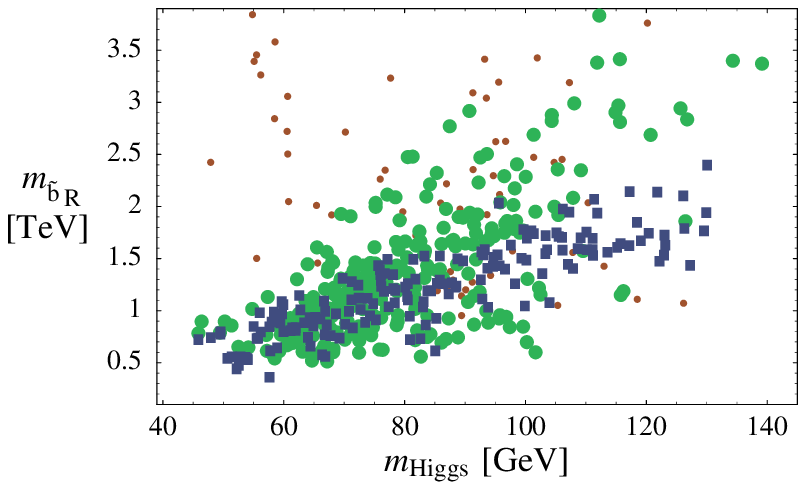,width=0.49\linewidth} \qquad
\epsfig{file=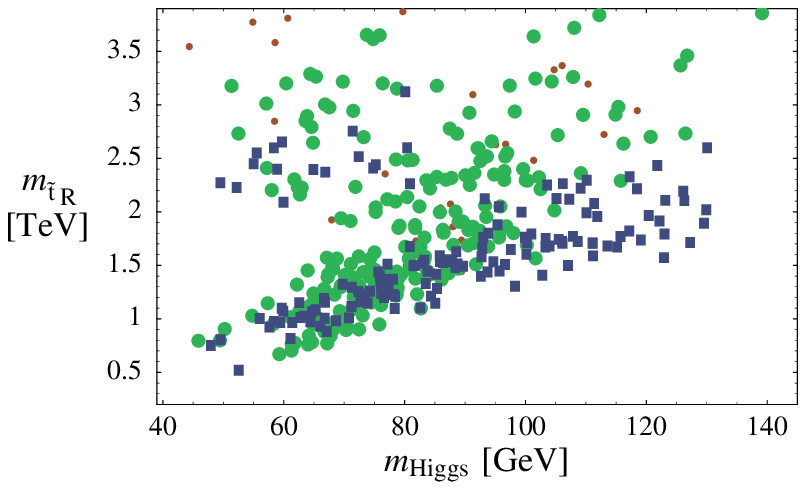,width=0.49\linewidth}
\end{minipage}
\caption{\it Masses of the first vectorial and fermionic resonances
$\rho$, $\tilde q_L$, $\tilde b_R$, $\tilde t_R$ 
obtained by scanning over the input parameters in the minimal model.
Blue squares correspond to $\eps>0.4$, 
green fat dots to $0.2<\eps<0.4$, small red dots to $\eps<0.2$.
}
\label{fig:mKK}
\end{figure}

Fig.~\ref{fig:standard} shows that the model can give moderately
small values of $\eps$, i.e.
$\eps\lesssim 0.4$, needed to pass all EWPT. This requires a
$10\%$ adjustment in the parameters of the model, as we already
explained in the previous section.
The value of $N$ can be large enough to guarantee a sensible perturbative
expansion, although for $m_\text{Higgs} \sim 115$ GeV it tends to
be small, $N\sim 4-5$, if one requires $\eps\sim 0.4$.
One of the most important predictions of the model is that the
Higgs mass is always light, $m_\text{Higgs}\lesssim 140$ GeV, for
$0.1 \lesssim \eps \leq 1$. 
This bound can be relaxed only for very small values of $\eps$,
due to the log enhanced top contribution 
$\propto \log(m_t/m_\rho) \sim \log \eps$.
However, a very small value of $\eps$ can be only obtained at
the price of a large fine tuning in the parameters of the model.
It should be stressed that the theoretical error on the
Higgs mass is expected to be large if $N$ is small. For example,
for $N\sim 5$ the correction to $m_\text{Higgs}$ coming from two
loop diagrams can be~$\sim 20\%$ or even larger.

The spectrum  of the lightest resonances, obtained by scanning
over the input parameter space, is shown in
figure~\ref{fig:mKK}. The fermionic states $\tilde q_L$, $\tilde t_R$, $\tilde b_R$
tend to be lighter than the gauge resonances. 
For values of $m_\text{Higgs} \sim
115$~GeV and $\eps\sim 0.4$, all fermionic resonances lie around $1.5-2$~TeV,
while $m_\rho\sim 2-3$~TeV.

\begin{figure}
\begin{minipage}[t]{0.95\linewidth}
\hspace{0.cm} \epsfig{file=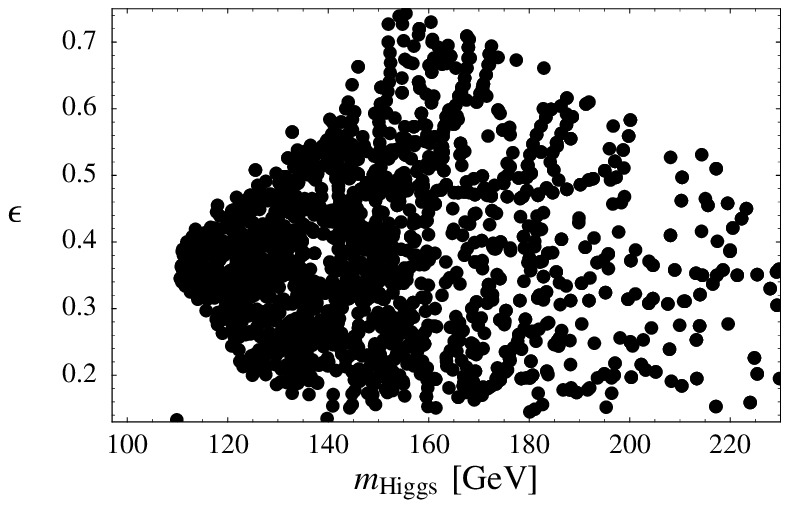,width=0.49\linewidth} \qquad
\epsfig{file=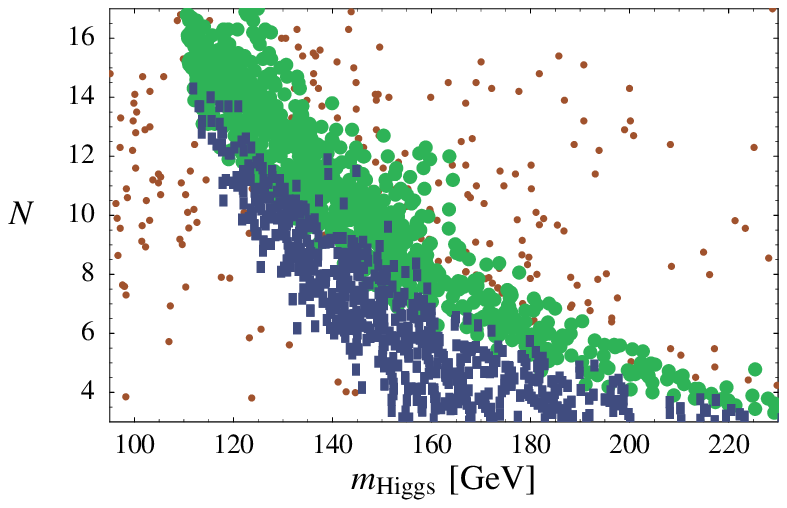,width=0.49\linewidth}
\end{minipage}
\caption{\it Scatter plots in the $(m_\text{Higgs},\eps)$ plane (left) and in the 
$(m_\text{Higgs},N)$ plane (right), obtained by scanning over the input parameters
with an extra contribution $\Delta V(h) = \xi\, \sin^4 h/f_\pi$, 
$\xi = -\beta_\text{gauge}$.
In the second plot, blue squares correspond to $\eps>0.4$, 
green fat dots to $0.2<\eps<0.4$, small red dots to $\eps<0.2$.
}
\label{fig:sin4}
\end{figure}

Finally, in fig.~\ref{fig:sin4} we show the results obtained by
adding a new contribution $\Delta V = \xi\, \sin^4 h/f_\pi$ to the Higgs
potential. The value of $\xi$ is
taken to be positive and equal in size to the gauge contribution
to $\beta$: $\xi = -\beta_\text{gauge} > 0$.
This example shows that small deformations of the Higgs potential
can help not only to eliminate the $10\%$ fine tuning 
of the minimal model, as discussed in sec.~\ref{vacmis},
but also to increase the range of the Higgs mass.
We checked that similar results also hold  taking
$\Delta V = \xi\, \sin^4 h/f_\pi + \zeta\, \sin^2 h/f_\pi$,
with $\zeta=-\xi=\beta_\text{gauge}$.
A possible origin of these new contributions is given in appendix~\ref{tree}.

\section{Third family EWPT: $Z\rightarrow b\bar b$ and the $T$  parameter}
\label{sec:EWPT}

\begin{figure}[t]
\centering
   \begin{picture}(120,80)
     \SetWidth{1} \GCirc(60,50){15}{0.85} \SetWidth{0.5}
     \ArrowLine(15,50)(45,50) \ArrowLine(75,50)(105,50) \SetWidth{1.5} \DashLine(60,35)(60,10){1.5}
     \SetWidth{0.5} \Text(25,60)[c]{$t_L$} \Text(95,60)[c]{ $t_R$} \Text(67,10)[lb]{$\Sigma$}
     \Text(60,-10)[c]{\large (a)}
   \end{picture} \hspace{0.8cm}
   \begin{picture}(120,80)
     \Photon(60,50)(60,10){2.5}{5}
     \ArrowLine(73.0,57.5)(92.9,69.0) \ArrowLine(27.1,69.0)(47.0,57.5)
     \SetWidth{1.5} \DashLine(60,50)(60,80){1.5}
     \SetWidth{1} \GCirc(60,50){15}{0.85} \SetWidth{0.5}
     \Text(67,10)[lb]{$Z$}
     \Text(24,82)[rt]{$b_L$} \Text(96,82)[lt]{$b_L$} \Text(60,89)[c]{$\Sigma$}
     \Text(60,-10)[c]{\large (b)}
   \end{picture} \hspace{1.1cm}
   \begin{picture}(120,80)
     \SetWidth{1.5} \DashLine(35,40)(20,14){1.5} \DashLine(35,40)(20,66){1.5}
                    \DashLine(85,40)(100,66){1.5} \DashLine(85,40)(100,14){1.5}
     \SetWidth{1} \GOval(35,40)(20,10)(0){0.85} \GOval(85,40)(20,10)(0){0.85} \SetWidth{0.5}
     \ArrowLine(35,60)(85,60) \ArrowLine(85,20)(35,20)
     \Photon(0,40)(25,40){2.5}{4} \Photon(95,40)(120,40){2.5}{4}
     \Text(0,50)[c]{$W_{\mu\nu}^{a_L}$}  \Text(120,50)[c]{$W_{\mu\nu}^{b_L}$}
     \Text(15,14)[r]{$\Sigma$} \Text(15,66)[r]{$\Sigma$}
     \Text(105,66)[l]{$\Sigma$} \Text(105,14)[l]{$\Sigma$}
     \Text(60,65)[b]{$t_R$} \Text(60,15)[t]{$t_R$}
     \Text(60,-10)[c]{\large (c)}
   \end{picture} \vspace{0.7cm}
\caption{\it Diagrams in the 4D holographic theory that generate
the top Yukawa coupling~(a), a correction to $Z\to b_L \bar b_L$~(b), 
and the $T$ parameter~(c). 
A grey blob represents the 4D CFT dynamics or the 5D bulk.
Another possible diagram contributing to $Z\to b_L \bar b_L$,
similar to (b) but with two $\Sigma$ fields attached, 
is not shown.} 
\label{fig:NDA}
\end{figure}
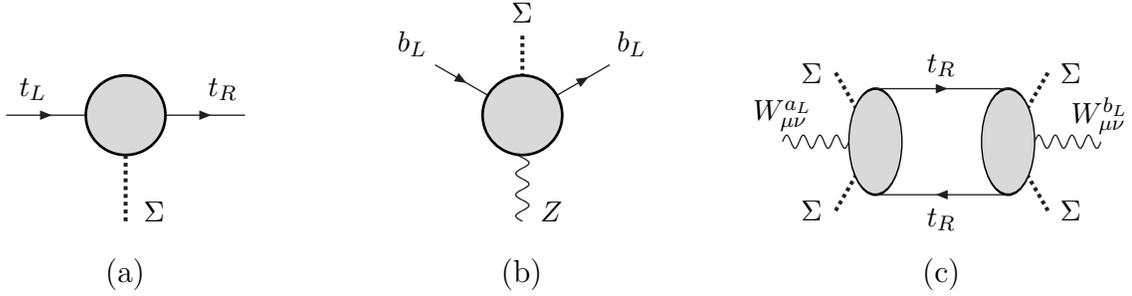

A large top Yukawa implies that the top
must substantially mix with the CFT (diagram fig.~\ref{fig:NDA}(a)).
This in turn suggests that there could be
large deviations from the SM prediction for $Z\rightarrow b\bar b$.
Such corrections are induced from the diagram of fig.~\ref{fig:NDA}(b).
Sizable one-loop contributions to $T$ are also expected
from fig.~\ref{fig:NDA}(c)
(see also refs.~\cite{Cacciapaglia:2004rb,Luty:2004ye} for a similar discussion).
In this section we will estimate these contributions
using NDA and show that they can be under control for $\eps\lesssim 0.4$.

Denoting with $\delta g_{Lb}$ the shift in the coupling of $b_L$ to $Z$,
NDA leads to the estimates~\footnote{If fig.~\ref{fig:NDA} is
drawn using resonances, one can show that there are  two kind of
diagrams contributing to $\delta g_{Lb}$ and $\Delta\rho$ . Either
the Higgs couples to a vector resonance, or to a fermionic resonance
through a chirality flip. One can show that the dominant
contribution to $\delta g_{Lb}$ and $\Delta\rho$ are respectively
that with  zero and four chirality flips.}
\begin{align}
\frac{\delta g_{Lb}}{g_{Lb}} &\sim \lambda^2_q \frac{ N }{ 16 \pi^2  }\, \eps^2
 \sim \left(\frac{1}{2}-c_q\right)\, \eps^2 \, ,
 \label{NDA1}\\
\alpha T  = \Delta\rho &\sim
 \lambda_u^4 \frac{ N }{ \left( 16 \pi^2 \right)^2 }\, \eta^4 \eps^2
 \sim  \left(\frac{1}{2}+c_u\right)^2 \eta^4\, \frac{\eps^2}{N}  \, ,
\label{NDA2}
\end{align}
where we have used eqs.~(\ref{lafp}) and (\ref{gac}). We have also
included a new parameter $\eta$ to take into account a possible
deviation from NDA in the coupling of the composite fermions to
the Higgs (a chirality flip factor).~\footnote{This corresponds in
the 5D theory to
 the mass mixing parameters eq.~(\ref{massmixing}).}
From eq.~(\ref{NDA1}) we see that the (liberal) bound 
$\delta g_{Lb}/g_{Lb}\lesssim 1\%$ from LEP is satisfied for values of $c_q$
close to 1/2. For example, $\eps\simeq 0.4$ implies $c_q\simeq 0.4$. 
From eq.~(\ref{NDA2}), on the other hand, we see that
the $99\%$ CL bound on the $T$ parameter, $T\lesssim
0.3$~\cite{Barbieri:2004qk}~\footnote{It corresponds to an extra
contribution to the $\eps_1$ parameter~\cite{eps123}
$\Delta\eps_1\lesssim 2.5 \cdot 10^{-3}$ relative to the SM value
with $m_\text{Higgs}$=115~GeV.}, can also be satisfied for
reasonable values of the parameters. For example, setting
$\eta\sim 1$, then $\eps\simeq 0.4$ implies $c_u\simeq -0.1$ for
$N\simeq 10$. Thus, both
estimates give $\delta g_{Lb}$ and $T$  close to the experimental
limit for values of  the 5D parameters
used in the analysis of section~\ref{sec:analysis}. This is an
indication that our model can succeed in passing all EWPT, 
although eqs.~(\ref{NDA1}),~(\ref{NDA2}) should not
be taken too seriously, being only estimates and not exact
results. 
One can take into account the correlation among $T$, $\delta g_{Lb}$
and the top mass by making use of the NDA estimate for $m_t$
\begin{equation} \label{NDA3}
m_t  \sim  \lambda_q \lambda_u
 \frac{N}{16\pi^2}\,  m_\rho\, \eps\,\eta
 \sim \sqrt{\left(\frac{1}{2}-c_q\right) \left(\frac{1}{2}+c_u\right)}\,
 \frac{4\pi}{\sqrt{N}}\, v\, \eta\ ,
\end{equation}
and combine it with eqs.~(\ref{NDA1}) and (\ref{NDA2}).
Imposing $\delta g_{Lb}/g_{Lb} \lesssim 1\%$ we then obtain
\begin{equation} \label{rhoepsilon}
\alpha T \gtrsim 0.4\, N\, \epsilon^6\, . 
\end{equation}
This is quite a stringent bound. Models with  $\eps =1$ give
$T\sim O(100)$ even for small $N$ and are therefore
grossly ruled out. Nevertheless, this bound on $T$ is extremely
sensitive to $\eps$, and  a value for $\eps$ slightly smaller than
1 can be  enough to satisfy the experimental constraint.

Our estimates (\ref{NDA1}), (\ref{NDA2}), (\ref{NDA3}) and
(\ref{rhoepsilon}) apply quite generally to a large class of 5D
models where fermions live in the bulk, EWSB is triggered in the
IR and custodial symmetry is violated only on the UV-brane.
The model of ref.~\cite{Agashe:2003zs},
where the Higgs is localized on the IR-brane, 
belongs to this class. In ref.~\cite{Agashe:2003zs} 
the $T$ parameter was explicitly
computed and found to be in agreement with eqs.~(\ref{NDA2}) and
(\ref{rhoepsilon}). The bound eq.~(\ref{rhoepsilon}) also applies
to 5D Higgsless models~\cite{higgsless,Barbieri:2003pr,Burdman:2003ya}. In that
case the 4D holographic description consists in a walking TC-like
theory where the two scales $f_\pi$ and $v$ coincide ($\eps=1$).
Thus, 5D Higgsless models are severely constrained by $T$. This is
true, we stress, even in the limit of strong bulk gauge coupling,
i.e. for a small value of  $N$.

We close this section mentioning a case in which
the bound~(\ref{rhoepsilon}) can be weakened.
This corresponds to the
 limit $c_u\to 1/2$ in which one can show that a massive $b^{ \prime }_R$
state of the CFT becomes light, and an extra factor
$(m_{b^{ \prime }}/m_\rho)^2$ suppresses the bound on $T$:
\begin{equation}
\alpha T \gtrsim 0.4\, N\,
 \frac{ m_{ b^{ \prime }}^2 }{ m_{ \rho }^2 }\, \epsilon^6\, .
\end{equation}
To understand this suppression factor it is convenient 
to adopt the holographic description in which the 
left-handed component  of  $\xi_u$,  
instead of the right-handed one,  
is fixed on the UV-brane.
In this case the holographic description is the 
following~\cite{Contino:2004vy}: the
right-handed top quark arises as a massless CFT bound state and,
by SU(2)$_R$ symmetry, it comes along with a massless partner
$b^\prime_R$. Together they form a doublet $Q_R=(t_R,b^\prime_R)$ of SU(2)$_R$.
An additional external field $b_L^{c\, \prime}$ exists, which marries $b_R^\prime$.
The external field $b_L^{c\, \prime}$
 represents the only source of violation
of the custodial symmetry. 
It is coupled to the CFT (and thus to $b_R^\prime$) through
a coupling which becomes marginal for $c_u=1/2$, and
irrelevant if $c_u > 1/2$.
In the limit $c_u\to +\infty$, $b_L^{c\, \prime}$
decouples from the CFT,  $m_{ b^{ \prime }}\to 0$,
the custodial symmetry is restored, and $T$ vanishes.
Therefore $T$ must be proportional to $m_{b^\prime}^2$.
The value of 
$m_{b^\prime}$, however, cannot be arbitrarily small, since
$b^\prime_R$ mixes also with  $b_L$ and induces an additional shift
in the coupling of $b_L$ to $Z$ of order $\delta g_{Lb}/g_{Lb} \sim
m_t^2 / m^2_{ b^\prime}$. 
By requiring $\delta g_{Lb}/g_{Lb} \lesssim 1\%$,
one has $m_{ b^\prime}\gtrsim 10\, m_t$.
Thus, $T$ can be somewhat reduced, but one still needs $\eps<1$ to be
consistent with the experimental constraint.

\section{Conclusions and comparison with other models of PGB Higgs}
\label{conclu}

Theories in which the Higgs arises as a PGB provide a rationale
for the smallness of the electroweak scale compared to the
scale of new physics, a fact that experiments seem to suggest.
This provides an important  motivation to
search for realistic theories of PGB Higgs.

Here we have presented the minimal  model in which
(\textit{i}) EWSB occurs dynamically, 
(\textit{ii}) the flavour problem is solved, and that
(\textit{iii}) is consistent with EWPT.
The Higgs appears as a composite PGB from a 5D bulk
with a symmetry breaking pattern SO(5)$\to$ SO(4).
EWSB occurs dynamically via top virtual effects,
and the Higgs mass can be calculated as a function of the 5D parameters
(see fig.~\ref{fig:standard}).

Other approaches to Higgs as PGB have been previously considered 
in the literature.
The Hosotani mechanism for EWSB~\cite{hosotani}, where a PGB Higgs
also appears as the fifth component of the 5D gauge field,
has been extensively studied~\cite{newhosotani}.
Nevertheless, none of these models is fully realistic.
The two important ingredients in our approach are the custodial symmetry and
that fact that we are working in a 5D  AdS space,
that  allows us to extrapolate the SM couplings
up to the Planck scale.~\footnote{Our model at low-energies 
$\simeq \mu_{\rm IR}$, however, could be realized
in flat space if  large UV-brane kinetic terms for the SM fields 
are added~\cite{Barbieri:2003pr}. 
What the AdS$_5$ geometry does is to give a dynamical
origin for these large UV-brane kinetic terms.}
As a consequence, we can  successfully address the flavour issue.
Furthermore, the AdS dynamics forces  us to  
work with a moderately large 5D gauge coupling, 
$g_5 \sqrt{k}\gtrsim 4$ (see eq.~(\ref{g5large})),
thus implying a large enough Higgs quartic coupling and  
a small $f_\pi/m_\rho$ as required by EWPT. 
This has to be compared with 5D flat theories with no
large brane kinetic terms.
In that case the 5D coupling is smaller,
$g_5 \sqrt{L^{-1}}\simeq g\simeq 0.65$
($L$ being the lenght of the fifth dimension),
and therefore the 1-loop Higgs quartic is too small.

Another approach to Higgs as a PGB has been
Little Higgs (LH) models~\cite{Arkani-Hamed:2001nc}.
In these models the electroweak scale is protected 
from the strong scale $\mu_{\rm IR}$,
up to  two-loop effects, due to collective breaking.
For this purpose, extra vector states $W^\prime$ are required.
The main benefit from  this approach is calculability. The theory
below $\mu_{\rm IR}$ is weakly coupled and therefore
predictable. This is different from our approach in
that we take a large $N$ limit in the strong sector
to obtain calculability.
Both approaches have a similar contribution to the $S$ parameter, since
the role of $m_\rho$ in our eq.~(\ref{rhos})
is now played by $m_{W^\prime}$.
In LH models, however,  the EW scale   is more sensitive
to $m_{W^\prime}$ than we are to $m_\rho$,
due to the fact that the one-loop contribution to the LH
potential is logarithmically divergent.
In our case the potential is finite and then less sensitive
to the scale of new physics.~\footnote{It is interesting to notice 
that in the limit of large kinetic terms on the IR-boundary,
i.e. small $g_{\rm IR}$,
our model resembles a LH model, since one resonance becomes lighter 
and more weakly coupled than the others (see eq.~(\ref{eq:scales})),
and plays the role of $W^\prime$.}

Phenomenologically  there are other important
differences. For example, differently from a LH theory,
in our model vector and fermion resonances come
in complete representations of SO(5).
The resonances $\tilde{q}_L$,  $\tilde{t}_R$  and $\tilde{b}_R$
are usually lighter than the vectorial ones (see figs.~(\ref{fig:mKK})),
and therefore their detection can be more feasible.
Their single production can proceed at the LHC via a Yukawa interaction:
\begin{equation}
\begin{gathered}
q_L + W_{long.}, \, Z_{ long.}\rightarrow
\tilde{t}_R, \tilde{b}_R\,  , \\
t_R + W_{ long. }, \, Z_{ long.} \rightarrow \tilde{q}_L\, ,
\end{gathered}
\end{equation}
where the incoming particles 
arise from the colliding protons.
Although the top quark content of the proton can be small,
the  production cross-section is enhanced by
the  large coupling involved in the vertex. 
This is because the couplings between CFT (5D bulk) states
are  of order $g_5\sqrt{k}\sim 4 $.
The decays of these KK fermions are just the inverse 
of their production process.
The $\tilde{t}_R$  production is similar to that of 
$t^{ \prime }$ in LH models. 
In refs.~\cite{Han:2003wu, Perelstein:2003wd} it has been shown
that $t^{ \prime }$ can be detected at the LHC if its mass is of 
order of few TeV. For $\tilde{q}_L$ and
$\tilde{b}_R$, however, there is no analog  in typical LH models.
The production of these fermionic resonances 
can also proceed via a virtual KK gluon~$\tilde g$ (with no analog in
LH models), predominantly for $\tilde{t}_R$:
\begin{equation}
q \bar{q} \rightarrow \tilde{g}^{ \ast }
 \rightarrow t_R + \tilde{t}_R\, .
\label{KKgluon}
\end{equation}
Here $q$ denotes a light or heavy quark. 
Finally, the gauge resonances can be 
produced via a Drell-Yan process, as in eq.~(\ref{KKgluon}).
For the case of   $W$ and $Z$ KK,
their production can also proceed by
$W_{ long. }$, $Z_{ long.}$ scattering 
(as in TC models)
due to the strong coupling between these states.
A detailed study of all these processes will be required 
to fully explore the phenomenology of our model at future colliders.

We think that the model presented here 
is a serious
alternative to supersymmetric models.
Several issues, however, must still be addressed. For example,
a full calculation of top quark effects on EWPT
($T$ parameter and $Z\to b\bar b$),
gauge coupling unification or string 
embedding.~\footnote{See ref.~\cite{Gubser:2004tf} for a first 
string realization of a PGB from an AdS background.}
We leave this for the future.

\newpage
\section*{Acknowledgments}

It is a pleasure to thank Nima Arkani-Hamed, Riccardo Barbieri, Paolo Creminelli,
Antonio Delgado, Gero von Gersdorff, David E. Kaplan, Markus Luty,
Frank Petriello, Riccardo Rattazzi, Claudio Scrucca and Alessandro Strumia 
for stimulating discussions.
We especially thank Raman Sundrum for collaboration in the early stages
of this project and for many discussions and comments.
The work of R.~C. is supported by NSF grants P420D3620414350 and P420D3620434350.
K.~A.~is supported by the Leon Madansky fellowship and NSF Grant P420D3620414350.

\appendix

\section{Fermionic self-energies}
\label{fselfen}

The fermionic self-energies $\Pi_i(p)$ and $M_i(p)$ of section~\ref{sec:effL} are easily
computed in the 5D theory by taking the
inverse of the propagator of the corresponding field with endpoints attached
to the UV-brane. A technical complication comes from the mixing among
different 5D bulk multiplets due to the IR boundary masses $\m$, $\M$
of eq.~(\ref{massmixing}). Resumming all possible insertions of the IR mass mixing
terms and omitting for simplicity boundary kinetic terms, we obtain
\begin{align}
\label{FFfirst}
\begin{split}
\Pi_{q_L}(p) =& \frac{k}{p^2}\, \frac{1}{G_{R\, q}^{(++)}(L_0,L_0)}
 \bigg\{
   1 - \frac{G_{R\, q}^{(++)}(L_0,L_1) G_{R\, q}^{(++)}(L_1,L_0)}{G_{R\, q}^{(++)}(L_0,L_0)}\times \\
   &\hspace{3.65cm} \times \, \frac{\m^2\, p^2\, (kL_1)^2\, G_{L\, q^u}^{(-+)}(L_1,L_1) }%
    {1-\m^2\, p^2\, (kL_1)^2\, G_{R\, q}^{(-+)}(L_1,L_1) G_{L\, q^u}^{(-+)}(L_1,L_1) }
 \bigg\} \, ,
\end{split}
\\[0.35cm]
\begin{split}
\Pi_{Q_L}(p) =& \frac{k}{p^2}\, \frac{1}{G_{R\, Q}^{(+-)}(L_0,L_0)}
 \bigg\{
   1 - \frac{\tilde G_{L\, Q}^{(-+)}(L_0,L_1)
             \tilde G_{R\, Q}^{(+-)}(L_1,L_0)}{G_{R\, Q}^{(+-)}(L_0,L_0)}\times \\
   &\hspace{3.65cm} \times \, \frac{\M^2\, (kL_1)^2\, G_{R\, Q^u}^{(++)}(L_1,L_1) }%
    {1-\M^2\, p^2\, (kL_1)^2\, G_{L\, Q}^{(++)}(L_1,L_1) G_{R\, Q^u}^{(++)}(L_1,L_1) }
 \bigg\} \, ,
\end{split}
\\[0.35cm]
\begin{split}
\Pi_{q^u_R}(p) =& \frac{k}{p^2}\, \frac{1}{G_{L\, q^u}^{(++)}(L_0,L_0)}
 \bigg\{
   1 - \frac{G_{L\, q^u}^{(++)}(L_0,L_1) G_{L\, q^u}^{(++)}(L_1,L_0)}{G_{L\, q^u}^{(++)}(L_0,L_0)}\times \\
   &\hspace{3.65cm} \times \, \frac{\m^2\, p^2\, (kL_1)^2\, G_{R\, q}^{(-+)}(L_1,L_1) }%
    {1-\m^2\, p^2\, (kL_1)^2\, G_{L\, q^u}^{(-+)}(L_1,L_1) G_{R\, q}^{(-+)}(L_1,L_1) }
 \bigg\} \, ,
\end{split}
\end{align}
\begin{align}
\begin{split}
\Pi_{Q^u_R}(p) =& \frac{k}{p^2}\, \frac{1}{G_{L\, Q^u}^{(+-)}(L_0,L_0)}
 \bigg\{
   1 - \frac{\tilde G_{R\, Q^u}^{(-+)}(L_0,L_1)
             \tilde G_{L\, Q^u}^{(+-)}(L_1,L_0)}{G_{L\, Q^u}^{(+-)}(L_0,L_0)}\times \\
   &\hspace{3.65cm} \times \, \frac{\M^2\, (kL_1)^2\, G_{L\, Q}^{(++)}(L_1,L_1) }%
    {1-\M^2\, p^2\, (kL_1)^2\, G_{R\, Q^u}^{(++)}(L_1,L_1) G_{L\, Q}^{(++)}(L_1,L_1) }
 \bigg\} \, ,
\end{split}
\\[0.35cm]
\begin{split}
M_q(p) =& -\m k^2 L_1\, \frac{G_{L\, q^u}^{(++)}(L_0,L_1)G_{R\, q}^{(++)}(L_1,L_0)}%
                          {G_{L\, q^u}^{(++)}(L_0,L_0)G_{R\, q}^{(++)}(L_0,L_0)} \times \\
 &\hspace{3.65cm} \times \frac{1}{1-\m^2\, p^2\, (kL_1)^2\,
                        G_{R\, q}^{(-+)}(L_1,L_1) G_{L\, q^u}^{(-+)}(L_1,L_1)}\, ,
\end{split}
\\[0.35cm]
\begin{split}
M_Q(p) =& -\frac{\M k^2 L_1}{p^2} \,
  \frac{\tilde G_{R\, Q^u}^{(-+)}(L_0,L_1) \tilde G_{R\, Q}^{(+-)}(L_1,L_0)}%
       {G_{L\, Q^u}^{(+-)}(L_0,L_0) G_{R\, Q}^{(+-)}(L_0,L_0)} \times \\
 &\hspace{3.65cm} \times \frac{1}{1-\M^2\, p^2\, (kL_1)^2\,
                        G_{L\, Q}^{(++)}(L_1,L_1) G_{R\, Q^u}^{(++)}(L_1,L_1)}\, .
\end{split}
\label{FFlast}
\end{align}
Here $G_{L,R}$ are defined as the left- and right-handed
components of the 5D propagator $S(p,z,z')$ of a bulk fermion with mass $c k$
between two points $z$, $z'$ along the fifth dimension 
(see for example~\cite{Contino:2003ve,Gherghetta:2000kr}):
\begin{equation}
S(p,z,z')= (k^2 z z')^{5/2}
 \left[ \pslash +\gamma^5 \left(\partial_z +\frac{1}{2z}\right)+\frac{c}{z} \right]
 \left[ P_R\, G_R(p,z,z')+P_L\, G_L(p,z,z') \right]\, ,
\end{equation}
where $P_{R,L}=(1\pm\gamma^5)/2$.
We have also defined
\begin{equation}
\tilde G_{R,L}(z,z') = \left[ \pm \partial_z + \frac{(c \pm 1/2)}{z} \right] G_{R,L}(z,z')\, .
\end{equation}

\section{A deformation of the Higgs potential}
\label{tree}

There are many possible sources of new contributions to the Higgs potential 
which can lead to the form of eq.~(\ref{sin4-sin2pot}).
For example, one could introduce two extra elementary fermions in the
4D theory, $\psi^{(T)}_L$ and $\psi^{(S)}_R$, 
the first transforming as a triplet under SU(2)$_L$, the latter as a singlet.
They can acquire a mass $\sim\mu_\text{IR}$ before EWSB by marrying some 
massless composite fermion, thus becoming heavy.
The leading one-loop potential generated by these extra fermions,
the analog of eq.~(\ref{apxpot}), is of the form 
$\Delta V=\tilde\alpha \sin^2 h/f_\pi + \tilde\beta \sin^4 h/f_\pi$.
This 4D modification of the minimal model can be obtained in the 5D
theory by adding two bulk fermions in the antisymmetric $\mathbf{10}$ 
representation of SO(5) with $(\pm,\mp)$ boundary conditions
for the various components.

In what follows we describe how the mechanism of ref.~\cite{Contino:2003ve}
can be applied to the present model to
generate a new contribution at tree level.
The idea is to introduce new scalar operators ${\cal O}_{\varphi_i}$
of dimension 2, coupled to external scalar sources $\varphi_i$
according to
\begin{equation}
\Delta{\cal L}=k\, \varphi_i{\cal O}_{\varphi_i}-\frac{1}{2}m^2_i \varphi^2_i\, ,
\end{equation}
with $k\sim M_\text{Pl}$.
The scalar sources do not need to be dynamical, but just
auxiliary fields that parametrize the breaking of the CFT global symmetry
($i$ labels the different components of an SO(5) representation).
Integrating out these scalars one obtains 
a marginal deformation of the CFT:
\begin{equation}
\Delta{\cal L}=
\lambda_i {\cal O}_{\varphi_i}^2\, ,
\label{deformation}
\end{equation}
where the coupling $\lambda_i$ runs logarithmically with energy.
At the scale $k$, $\lambda_i(k)=k^2/(2m^2_i)\sim 1$ for $m_i\sim k$.
This SO(5)-breaking contribution will generate 
new terms in the PBG potential.

Let us consider for example the case in which the fields 
$\varphi_i$ fit into a traceless symmetric representation of SO(5), 
a ${\bf 14_0}$ of SO(5)$\times$U(1)$_{B-L}$. 
A \textbf{14} decomposes as
\textbf{14}=\textbf{9}+\textbf{4}+\textbf{1} 
under SO(4), where \textbf{9}=(\textbf{3},\textbf{3}),
\textbf{4}=(\textbf{2},\textbf{2}), \textbf{1}=(\textbf{1},\textbf{1}) 
under SU(2)$_L\times$SU(2)$_R$.
The different components 
can have different mass terms, $m_i^2$  ($i=9,4,1$),
because SO(5) is not a symmetry of the external sector.
We are considering, for simplicity, the case in which 
these masses respect an SO(4) symmetry.
Integrating out the scalars at tree-level, one obtains the 
following contributions to the Higgs potential:
\begin{equation}
\label{potextra}
\Delta V(\Sigma) = -\frac{\lambda_\varphi^2}{2} m^4_\rho\left[
\frac{1}{m_1^2} (\Sigma \hat T^0 \Sigma^T)^2
+  \frac{1}{m_4^2} (\Sigma \hat T^a \Sigma^T)^2
+ \frac{1}{m_9^2}  (\Sigma \hat T^n \Sigma^T)^2 \right]\, ,
\end{equation}
where $\hat T^{A=0,a,n}$ are a set of traceless symmetric matrices given in
 appendix~\ref{generators}.
The coupling $\lambda_\varphi$ runs logarithmically with energy.
Using eq.~(\ref{eq:Sigma}) in eq.~(\ref{potextra}) one has:
\begin{equation} \label{scalarpot}
\Delta V(h) = - \lambda_\varphi^2 m^4_\rho \left[
  \left(\frac{5}{8}\frac{1}{m_1^2}- \frac{1}{m_4^2}
 +\frac{3}{8}\frac{1}{m_9^2} \right) \sin^4 \frac{h}{f_\pi}
 +\left(\frac{1}{m_4^2}-\frac{1}{m_1^2}\right) \sin^2 \frac{h}{f_\pi}
 +\frac{2}{5}\frac{1}{m_1^2}\right]\, .
\end{equation}
The overall coefficient $\lambda_\varphi$ can be easily of loop size if
the coupling of $\varphi_i$ to the CFT is not exactly maximally relevant.
It is possible to generate only a  $\sin^4h/f_\pi$  term
if the $\textbf{9}$ is the only external source present ($m_{1,4}\to\infty$).
To have a positive coefficient, $m_9^2$ must be negative,
which implies that the $\textbf{9}$ should not propagate 
(otherwise we have a  tachyon).
It must be  considered  an auxiliary field,
like a D-term in supersymmetry.
Another possibility is to assume $m_4 \simeq m_1$ as the result of some
symmetry of the UV physics.
The other particular form of the extra contribution to the potential
mentioned in the text, $\Delta V = \xi\, (\sin^4 h/f_\pi - \sin^2 h/f_\pi)$,
can be obtained if the $\textbf{4}$ is the only external source present 
($m_{1,9}\to\infty$).

The 5D realization of this 4D mechanism follows straightforwardly
through holography~\cite{Contino:2003ve}. An additional scalar
field $\Phi$ of 5D mass $M_\Phi^2 = -4k^2$ lives in the bulk and
transforms as a $\textbf{14}$ of SO(5). If its singlet component
under SO(4) has a tadpole on the IR-brane, the potential
(\ref{scalarpot}) will be generated at tree level. Finally,
dynamical (non-dynamical) fields of the 4D external sector, $\varphi_i$,
correspond to the components of $\Phi$ on the UV-brane 
with a Neumann (Dirichlet)
boundary condition.

\section{SO(5) generators}
\label{generators}

We collect here the SO(5) generators used in the text. A suitable basis
for the vectorial representation is
\begin{equation}
\begin{split}
T^{a_{L,R}}_{ij} =& -\frac{i}{2}
 \left[ \frac{1}{2}\, \eps^{abc} \left(\delta^b_i \delta^c_j - \delta^b_j \delta^c_i \right)
        \pm \left(\delta^a_i \delta^4_j-\delta^a_j \delta^4_i \right) \right]\\
T^{\hat a}_{ij}  =& -\frac{i}{\sqrt{2}}
 \left( \delta^{\hat a}_i \delta^5_j - \delta^{\hat a}_j \delta^5_i \right)
\end{split}
\end{equation}
where $i,j=1,\dots,5$ and  $T^{\hat a}$ ($\hat a=1,\dots,4$), $T^{a_{L,R}}$ ($a_{L,R}=1,2,3$)
are respectively the generators of SO(5)/SO(4) and SO(4)$\sim$SU(2)$_L\times$SU(2)$_R$.
The spinorial representation of SO(5) can be defined in terms of the Gamma matrices
\begin{equation}
\Gamma^{\hat a} =
 \begin{bmatrix} 0 & \sigma^{\hat a} \\ \sigma^{\hat a\,\dagger} & 0 \end{bmatrix}\, , \qquad
\Gamma^5 = \begin{bmatrix} \mathbf{1} & 0 \\ 0 & -\mathbf{1} \end{bmatrix}\, ,
 \qquad\qquad \sigma^{\hat a}=\{\vec\sigma,-i \mathbf{1}\} \, ,
\end{equation}
as
\begin{equation}
T^{a_{L,R}} = -\frac{i}{2\sqrt{2}}
 \left[ \frac{1}{2}\, \eps^{abc} [\Gamma^b,\Gamma^c] \pm [\Gamma^a,\Gamma^4] \right]\, , \qquad
T^{\hat a}  = -\frac{i}{4\sqrt{2}} [\Gamma^{\hat a},\Gamma^5]\, ,
\end{equation}
so that
\begin{equation}
T^{a_L} = \frac{1}{2} \begin{bmatrix} \sigma^a & 0\\ 0 & 0 \end{bmatrix}\, , \qquad
T^{a_R} = \frac{1}{2} \begin{bmatrix} 0 & 0\\ 0 & \sigma^a \end{bmatrix}\, , \qquad
T^{\hat a} = \frac{i}{2\sqrt{2}}
 \begin{bmatrix} 0 & \sigma^{\hat a} \\ -\sigma^{\hat a\, \dagger} & 0 \end{bmatrix}\, .
\end{equation}
Finally, we give the expression of the orthogonal basis of traceless symmetric
5$\times$5 matrices used in the previous appendix. Denoting with
$\hat T^0$, $\hat T^a$ ($a=1,\dots,4$), $\hat T^n$ ($n=1,\dots,9$) respectively the
\textbf{1}, \textbf{4} and \textbf{9} representations of the SO(4) subgroup, one has:
 \begin{equation}
 \begin{gathered}
 \hat T^0_{ij} = \frac{1}{2\sqrt{5}}\, \text{diag}(1,1,1,1,-4)\, , \qquad
 \hat T^a_{ij} = \frac{1}{\sqrt{2}}
  \left( \delta^a_i \delta^5_j + \delta^a_j \delta^5_i\right)\, ,  \\
 \hat T^n_{ij} = \Big\{ t^{bc}_{ij}\, ;
  \frac{1}{\sqrt{2}}\, \text{diag}(-1,0,1,0,0)\, ;
  \frac{1}{2\sqrt{3}}\, \text{diag}(1,1,1,-3,0)\, ;
  \frac{1}{\sqrt{6}}\, \text{diag}(-1,2,-1,0,0)  \Big\}\, ,
 \end{gathered}
 \end{equation}
where $t^{bc}_{ij}= (\delta^b_i \delta^c_j + \delta^b_j \delta^c_i)/\sqrt{2}\; , c>b$
$(b,c=1,\dots,4)$ is a collection of six matrices.


\end{document}